\documentstyle[12pt]{article}

\begin{document}

\bigskip\bigskip

\begin{center}
{\Large {\bf The Vacuum Energy from a New Perspective}}

\bigskip\medskip

{\large Antonio R. Mondragon and Roland E. Allen}

\medskip

{\small {\it Center for Theoretical Physics, Texas A\&M University, \\[0 pt]
College Station, Texas 77843, USA }}

\bigskip\medskip

\parbox{5.4125 in} {\small {\bf Abstract.} It is commonly believed 
that the vacuum energy problem points to the need for 
(1) a radically new formulation 
of gravitational physics and (2) a new principle which forces the 
vacuum stress-energy tensor (as 
measured by gravity) to be nearly zero. Here we point out that 
a new fundamental theory contains both features: (1) In this theory 
the vierbein is interpreted as the ``superfluid velocity'' associated 
with the order parameter $\Psi_{s}$ for a GUT-scale Higgs condensate. 
(2) The vacuum stress-energy tensor ${\cal T}^{vac}_{\mu \nu }$ is 
exactly zero in the vacuum state, because the 
action is extremalized with respect to variations in $\Psi_{s}$. With 
inhomogeneously-distributed matter present, ${\cal T}^{vac}_{\mu \nu }$ 
is shifted away from zero.}
\end{center}

\bigskip
The vacuum energy is one of the deepest issues in theoretical physics, and
no conventional theory -- including superstring/M theory -- has offered a
convincing solution to the problem of why the vacuum stress-energy 
tensor (as measured by gravity) is vastly 
smaller than expected but still nonzero~[1-3].

In this paper we consider an unconventional theory which contains a
radically new formulation of gravitational physics~[4-6]. The gravitational
vierbein is interpreted as the ``superfluid velocity'' of a GUT-scale
condensate $\Psi _{s}~$ which forms in the very early universe:
\begin{equation}
g^{\mu \nu }=\eta ^{\alpha \beta }e_{\alpha }^{\mu }\,e_{\beta }^{\nu }
\quad,\quad e_{\alpha }^{\mu }~=~v_{\alpha }^{\mu } 
\quad ,\quad v^{\mu }=v_{\alpha }^{\mu }\,
\sigma ^{\alpha }~~\mbox{with}~~\mu ,\alpha=0,1,2,3
\end{equation}
\begin{equation}
v^{\mu }=\eta ^{\mu \nu }v_{\nu }
\quad ,\quad mv_{\mu }~=~-iU^{-1}\partial_{\mu }U
\quad ,\quad \Psi _{s}=n_{s}^{1/2}U\eta _{s}
\quad ,\quad \eta_{s}^{\dagger }\eta _{s}=1.
\end{equation}
Here $\eta ^{\mu \nu }$ = diag (-1,1,1,1) is the Minkowski metric tensor,
the $\sigma ^{\alpha }$ are the identity matrix and three Pauli matrices, 
$U$ is a $2 \times 2$ unitary matrix, 
$\eta _{s}$ is a constant 2-component vector, 
and $n_{s}$ is the condensate density. (After a Kaluza-Klein reduction from
a higher-dimensional theory, the initial group of this order parameter is 
$SO(10)\times SU(2)\times U(1)$. For the purposes of this paper, however, the
gauge group $SO(10)$ can be ignored, leaving the simpler description of
(1) and (2).) In the present theory, $\Psi _{s}$ is not static but instead
exhibits $SU(2)\times U(1)$ rotations as a function of position and
time. This condensate also supports Planck-scale $SU(2)$ instantons 
(in a Euclidean picture), which 
are analogous to the $U(1)$ vortices in an ordinary superfluid. In the
present theory, the Einstein-Hilbert action and the curvature of spacetime
result from these instantons~\cite{allen1}. Quantum gravity has a natural
cutoff at the energy scale $m\sim $ the Planck energy $m_{P}$, but at lower
energies one recovers the Einstein field equations 
\begin{equation}
\frac{\delta S_{total}}{\delta g^{\mu \nu }}=\frac{\delta S_{vac}}{\delta
g^{\mu \nu }}+\frac{\delta S_{fields}}{\delta g^{\mu \nu }}+\frac{\delta
S_{EH}}{\delta g^{\mu \nu }}=0
\end{equation}
or 
\begin{equation}
R_{\mu \nu }-{\frac{1}{2}}g_{\mu \nu }{}^{(4)}R= 
{\frac{1}{2}} \ell_{P}^{2} {\cal T}_{\mu \nu }^{total}=
{\frac{1}{2}} \ell_{P}^{2} \left( {\cal T}_{\mu \nu }^{vac}+
{\cal T}_{\mu \nu }^{fields} \right)
\end{equation}
where 
\begin{equation}
{\cal T}_{\mu \nu }^{vac} = -\frac{2}{\sqrt{-g}}\frac{\delta S_{vac}}
{\delta g^{\mu \nu }}\quad ,\quad {\cal T}_{\mu \nu }^{fields} = -
\frac{2}{\sqrt{-g}}\frac{\delta S_{fields}}{\delta g^{\mu \nu }}
\end{equation}
and $\ell_{P}$ is the Planck length defined in Ref. 4.

In the vacuum state, there is no stress-energy tensor for matter 
and radiation: ${\cal T}_{\mu \nu }^{fields}=0$. We additionally assume that 
there there is no contribution from topological defects in the vacuum 
state: $\delta S_{EH}/\delta g^{\mu \nu }=0$. This  
assumption will be discussed in more detail elsewhere~\cite{allen3}, 
but it will be seen below that it leads to a consistent solution (whereas 
such a solution could not be obtained in conventional physics). 
We then have 
\begin{equation}
\delta S_{vac}=\delta S_{total}=0\quad \mbox{in the vacuum state}
\end{equation}
for arbitrary variations of the order parameter $\Psi _{s}$. $~$Variations in 
$v_{\alpha }^{\mu }$, however, are a special case of functional variations 
in $\Psi_{s} $. 
It follows that the vacuum stress-energy tensor is exactly zero: 
\begin{equation}
{\cal T}_{\mu \nu }^{vac}=0\quad \mbox{in the vacuum state}.
\end{equation}

It is interesting to see in more detail how (7) can be achieved.
According to the quantum Bernoulli equation (3.20) of Ref. 4, we have 
\begin{equation}
{\frac{1}{2}}m\eta _{s}^{\dagger }\eta _{\mu \nu }v^{\mu }v^{\nu }\eta_{s}
+ V  + P + V_{vac} = \mu  
\end{equation}
\begin{equation}
V = b n_{s} \quad ,\quad 
P = - \frac{1}{2m}n_{s}^{-1/2}\eta ^{\mu \nu }\partial _{\mu } 
\partial _{\nu }n_{s}^{1/2}
\end{equation}
where $\mu $ is a fundamental energy which 
plays the role of a chemical potential 
here and which is comparable to $m_{P}$. We have added a term 
$V_{vac}$ which represents the contribution of all other vacuum 
fields to $\delta S_{vac}$ when $\Psi_{s}$ is varied. Let us rewrite 
(8) as
\begin{equation}
- \frac{1}{2m}n_{s}^{-1/2} \eta ^{\mu \nu } 
\partial _{\mu}\partial _{\nu }n_{s}^{1/2}+ b n_{s} 
= \mu - V_{vac} - {\frac{1}{2}}m\eta_{\mu \nu }e^{\mu}_{\alpha}e^{\nu }_{\beta}
\eta _{s}^{\dagger } \sigma^{\alpha} \sigma^{\beta} \eta_{s}.
\end{equation}
After $V_{vac}$ and $e^{\mu}_{\alpha}$ are specified, the 
condensate density $n_{s}$ adjusts itself to satisfy (10), (6), and (7). 
One might express this result as follows: In the vacuum 
state, the vacuum stress-energy tensor is tuned to exactly zero 
through adjustments of the condensate density. Notice that 
the extremalization (3) in conventional physics requires a 
contribution from the Einstein-Hilbert action 
$S_{EH}$ even when $\delta S_{fields}/\delta g^{\mu \nu }=0$, but in 
the present theory this extremalization in the vacuum state can be 
accomplished with $S_{vac}$ alone.

\pagebreak

In a more general state with matter, radiation, and topological defects 
present, it is the {\it total} action which is extremalized in (3). Then 
$\delta S_{vac}/\delta g^{\mu \nu }$ is shifted away from zero: 
\begin{equation}
{\cal T}_{\mu \nu }^{vac}\neq 0\quad \mbox{with matter and radiation present}.
\end{equation}

There are two primary aspects of the vacuum energy 
problem~\cite{weinberg1,weinberg2}: 
(i) Why is the vacuum stress-energy tensor many orders of magnitude smaller 
than predicted by conventional physics? This question is addressed in (7). 
(ii) Why is the vacuum stress-energy tensor not exactly zero? 
This is addressed in (11).  

Although these are the ``big'' questions, one can add two more in the
present context: (iii) Why was the vacuum energy density small compared to
the density of matter and radiation during the period of big-bang
nucleosynthesis? (iv) Why is it comparable to the density of matter now?
To fully answer these questions will require a detailed treatment of how the
vacuum energy is affected by the presence of matter and radiation. Suppose,
however, that the dominant mechanism is a Casimir-like effect, in which the
vacuum energy is modified by the boundary conditions imposed on the vacuum
fields when there is an inhomogeneous distribution of matter. In a
radiation-dominated universe, the energy density will be relatively
homogeneous, and such an effect should be relatively small. 
In the present epoch, 
on the other hand, there is an extremely inhomogeneous distribution of matter.
This plausibility argument indicates that the vacuum energy should play an
important role only in the present epoch, and that the vacuum energy density
(as measured by the stress-energy tensor) will be comparable to the density
of inhomogeneously-distributed matter.

\bigskip\bigskip {\Large {\bf Acknowledgement}}

\bigskip This work was supported by the Robert A. Welch Foundation.


\begin{thebibliography}{9}
\bibitem{weinberg1} S. Weinberg, {\it Rev. Mod. Phys.} 61, 1 (1989).

\bibitem{weinberg2} S. Weinberg, talk at the 4th International Symposium 
on Sources and Detection of Dark Matter in the Universe (DM2000) 
and astro-ph/0005265.

\bibitem{witten}  E. Witten, talk at DM2000 and hep-ph/0002297.

\bibitem{allen1} R. E. Allen, {\it Int. J. Mod. Phys. A}
12, 2385 (1997) and hep-th/9612041.

\bibitem{allen2}  R. E. Allen, to be published and
hep-th/0008032.

\bibitem{allen3}  R. E. Allen, to be published and talk
given at the conference on Problems with Vacuum Energy (Copenhagen, August
24-26, 2000).

\end{thebibliography}
\end{document}